\documentclass[11pt]{article}
\usepackage{amsmath,amsfonts,amsthm,amssymb,epsf,color}

\newcommand{\eq}{\begin{equation}}
\newcommand{\en}{\end{equation}}
\newcommand{\eqa}{\begin{eqnarray}}
\newcommand{\ena}{\end{eqnarray}}

\begin{document}

 \setlength{\unitlength}{1mm}

 \vspace*{0.1cm}


\begin{center}

   {\bf Quantum Computing via The Bethe Ansatz}
 \vspace{.3cm}

 Yong Zhang\footnote{zhangyo2008@gmail.com}\\ \vspace{.2cm}

 \em{Zhuopu Zhonghu Computing Company, Wuhan 430080, P.R.China\\ }

\end{center}

\vspace{0.2cm}

 \begin{center}

 \parbox{12.1cm}{
 \hspace{18pt} We  recognize quantum circuit model of computation as factorisable
 scattering model and propose that a quantum computer is associated with a
 quantum many-body system solved by the Bethe ansatz. As an typical example to support
 our perspectives on quantum computation, we study quantum computing in one-dimensional
 nonrelativistic system with delta-function interaction, where the two-body scattering
 matrix satisfies the factorisation equation (the quantum Yang--Baxter equation) and acts
 as a parametric two-body quantum gate.  We conclude by comparing quantum computing via 
 the factorisable scattering with topological quantum computing.
}

\end{center}

\vspace{1.cm}


 Computers are physical objects, and computations are physical processes. This sentence
 becomes one of Deutsch's famous quotes, and is a helpful guiding principle to understand
 Deutsch's original perspectives on quantum computation in his two seminal papers
 \cite{Deutsch85, Deutsch89}. With it, {\em a large-scale quantum computer is a quantum
 many-body system, and its motion from an initial state to a final state is  the performance
 of quantum computation from an input to an output}. An explicit example for this
 statement is the quantum circuit model of computation \cite{Deutsch89}, which expresses
 an output as the result of a sequence of quantum gates on an input, in which a quantum
 gate is a unitary transformation with its input and output and implemented by a quantum
 process with the initial state and the final state.

 An arbitrary $N$-qubit unitary quantum gate has the following factorization property.
 It can be expressed exactly as a sequence of products of some two-qubit gates
 \cite{Barenco95a}, which are generated by the CNOT gate
 with one-qubit gates \cite{Barenco95b}. Hence a quantum circuit consists of only
 two-qubit gates or two-qubit gates with one-qubit gates.
 On the other hand, Deutsch \cite{Deutsch89} denoted an $N$-qubit quantum gate by a
 $2^N\times 2^N$ scattering matrix (S-matrix) since both 
 are unitary matrices with the equal number of inputs and outputs, and intuitively viewed
 quantum computation performed by an $N$-qubit quantum gate as an $N$-qubit quantum
 elastic scattering process. Therefore, {\em a quantum circuit model can be viewed as
 a factorisable scattering model} \cite{Sutherland04}, which is our observation on the
 physics underlying the quantum circuit model.

 Suppose that we already have a quantum computer able to perform universal quantum
 computation or a specified class of computation tasks. To set up and operate on such
 a machine,  it is the best to solve the Schr{\"o}dinger equation of this many-body system
 to derive its exact wavefunction and then have its all key physical properties. That is,
  {\em a quantum computer is associated with an exactly solvable quantum
 many-body system}, also because many known exactly solvable models are integrable models \cite{Sutherland04}
 naturally giving rise to {\em factorisable scattering models}.

 Almost all known integrable models are solved by the Bethe ansatz \cite{Sutherland04}.
 The ansatz is an assumption on the wavefunction of a quantum many-body system and is able to transform
 the problem of solving the Schr\"odinger equation into the problem of solving algebraic equations.
 It was originally proposed by Bethe in the study of the Heisenberg spin chain, and later was
 used by many authors to solve one-dimensional model with $\delta$-function interaction,
 one-dimensional Hubbard model, etc.  Our proposal can be hence stated
 as follows, {\em a quantum circuit model is associated with a quantum many-body system solved by the
 Bethe ansatz}.  That is, we choose to explore physics of quantum computer with the Bethe
 ansatz. In Sutherland's book \cite{Sutherland04}, the asymptotic Bethe ansatz has the same content
 as the {\em factorisable scattering}, and so equivalently, we focus on quantum computing via the
 {\em factorisable scattering}.

 Many-body quantum scattering in two dimensional space-time is called {\em factorisable scattering}
 \cite{McGuire64} if and only if it has the properties: all collisions are purely elastic processes;
 only two-body collisions are allowed; the order of two-body collisions can be adjusted under the
 consistency condition.  Hence the $N$-body  factorisable collision is  expressed as a sequence of
 two-body collisions, i.e., the $N$-body factorisable scattering matrix ($S$-matrix) is equal to a
 product of $N(N-1)/2$ two-body $S$-matrices. The two-body $S$-matrix satisfies the consistence
 condition called the factorization equation, which was firstly discovered by  McGuire \cite{McGuire64}
 and is now often referred to as the quantum Yang--Baxter equation \cite{Yang67}.

 Nontrivial unitary solutions of the factorization equation, i.e., two-body scattering matrices,
 have been recognized by the author and his coauthors \cite{ZKG05a, ZKW07, Zhang07} as parametric
 two-body quantum gates from the viewpoint of mathematicians or mathematical physicists.  The author
 named {\em integrable quantum computing} to stand for quantum computing via non-trivial unitary solutions of
 the factorisation equation in his previous paper \cite{Zhang08}, which can be generalized to represent
 {\em quantum computing via the Bethe ansatz}, i.e., the title of this paper.  As an example, let us
 focus on {\em integrable quantum computing} in one-dimensional  nonrelativistic system with
 $\delta$-function interaction, and interested readers are invited to refer to Bose and Korepin's
 latest preprint \cite{BK11a} as a preliminary introduction on how to have the two-body scattering
 matrix as a quantum gate between two flying particles.

 Take Yang's notation \cite{Yang67} for the two-body scattering matrix,
 \eq
  Y^{12}_{12}=\frac {i(p_1-p_2) P_{12} +c } {i (p_1-p_2)-c} \nonumber
 \en
where $c$ denotes interaction strength and $c>0$ ($c<0$) means repulsive (attraction) interaction,
$p_1$ and $p_2$ represent momenta of two flying particles respectively, and $P_{12}$ denotes the permutation
operator (the Swap gate) acting on internal states of these two flying particles
\footnote{The permutation operator $P_{12}$ is noted for identical bosonic particles or distinguishable
particles, while $-P_{12}$ is used for identical fermionic particles. Bose and Korepin's notation on the
two-body scattering matrix \cite{BK11a}  is obtained from ours by multiplying the permutation operator
$P_{12}$ and exchanging the positions of $p_1$ and $p_2$.}.
Note that the two-body scattering matrix in the Heisenberg spin chain and
the Hubbard model \cite{Sutherland04} has a similar form of  $Y^{12}_{12}$, and so our research
can be applied to these models in principle.

Define a new variable $\varphi$ in the way,
\eq
\tan \varphi = \frac {p_2 - p_1} c, \quad\quad -\frac \pi 2 < \varphi < \frac \pi 2, \nonumber
\en
and it gives rise to a new form of the scattering matrix $Y^{12}_{12}$ by
\eq
Y^{12}_{12}(\varphi)=-e^{-2 i \varphi} \left(\begin{array}{cccc} 1 & 0 & 0 & 0 \\
   0&   \frac {1+e^{2 i\varphi}} 2 & \frac {1-e^{2i\varphi}} 2 & 0 \\
    0 & \frac {1-e^{2i\varphi}} 2 &  \frac {1+e^{2 i\varphi}} 2 & 0\\
     0 & 0 & 0 & 1   \end{array}  \right), \nonumber
\en
which can be implemented as a parametric two-body gate modulo a phase factor in the scattering
process between a flying qubit and a static qubit \cite{CCOZCB10}. As the parameter $\varphi$ has the
value $\pm \frac \pi 4$, i.e., $p_2-p_1=\pm c$, we have a maximal entangling gate to yield
the Bell states with local unitary transformations.

Introduce another symbol $\alpha=i \tan \varphi$, the scattering matrix $Y^{12}_{12}$ has the form
\eq
Y^{12}_{12}=-\frac 1 {1+\alpha} (1-\alpha P_{12}) \nonumber
\en
which is a rational solution of the factorisation equation and has been associated with the Werner state,
see the author and his coauthors' paper \cite{ZKW07} for the detail. Therefore, the entangling properties
of the scattering matrix $Y^{12}_{12}$ can be explored with the help of the entangling properties of the
Werner state. In addition, the Werner state is invariant under $SU(2)\times SU(2)$, and so the scattering
matrix $Y^{12}_{12}$ does. This too much symmetry decides the scattering matrix $Y^{12}_{12}$ not to be a
universal quantum gate \cite{Barenco95a} generating all unitary transformations.

In terms of $\varphi$, the scattering matrix $Y^{12}_{12}$ has an exponential form of $P_{12}$,
\eq
Y^{12}_{12}(\varphi)=e^{i(\pi-\varphi)} e^{-i\varphi P_{12}}, \quad -\frac \pi 2 < \varphi < \frac \pi 2 \nonumber
\en
which suggests that it is impossible to have an exact Swap gate $P_{12}$ directly by taking
$\varphi=\frac \pi 2$ but possible to approximate the Swap gate by taking $p_2-p_1 \gg c$, i.e.,
$\varphi=\frac \pi 2 + \epsilon$ with infinitesimal parameter $\epsilon$ in the way
\eq
Y^{12}_{12}(\frac \pi 2 +\epsilon) =P_{12} +{\mathcal O}(\epsilon). \nonumber
\en
Fortunately, there exists the other way to have an exact Swap gate modulo a phase factor by use of
the $\sqrt{Swap}$ gate. As $p_2-p_1=c$, the scattering matrix $Y^{12}_{12}$ has the form
\eq
Y^{12}_{12}(\frac \pi 4)= i \sqrt{Swap},\nonumber
\en
with the $\sqrt{Swap}$ gate given by
\eq
\sqrt{Swap} = P_{+} + i P_{-}, \quad P_{+}=\frac {1+P_{12}} 2, \quad P_{-}=\frac {1-P_{12}} 2,\nonumber
\en
which implements the Swap gate modulo a phase factor in the way
\eq P_{12}=-Y^{12}_{12}(\frac \pi 4) \cdot Y^{12}_{12}(\frac \pi 4).\nonumber \en
As Loss and DiVincenzo \cite{LD98} pointed out: two $\sqrt{Swap}$ gates with three one-qubit
gates yield the CNOT gate, and hence we can perform universal quantum computation via the scattering 
matrix $Y^{12}_{12}(\frac \pi 4)$ with local unitary transformations on flying qubits.

In terms of the Heisenberg interaction $\vec{S}_1\cdot \vec{S}_2$
between two spin-$1/2$ particles, the Swap operator $P_{12}$ has a form
\eq
 P_{12} =2 \vec{S}_1\cdot\vec{S}_2 + \frac 1 2,\nonumber
\en
and the two-body scattering matrix $Y_{12}^{12}(\varphi)$ can be expressed as a time evolution
$U_{12}(\varphi)$ of the Heisenberg interaction modulo a phase factor,
\eq
 Y^{12}_{12}(\varphi)=-e^{-i\frac 3 2 \varphi} U_{12}(2\varphi),
  \quad U_{12}(\varphi)=e^{-i\varphi \vec{S}_1\cdot\vec{S}_2} \nonumber
\en
which means that $Y^{12}_{12}$ and $U_{12}$ are two equivalent parametric two-body quantum
gates in the quantum circuit model. Obviously $U_{12}$ is not a universal two-qubit gate \cite{Barenco95a},
but DiVincenzo et al.,  \cite{DiVincenzo00}  can set up universal quantum computation only using the
exchange interaction $U_{12}$ if a qubit is encoded as a two-dimensional subspace of eight-dimensional
Hilbert space of three spin-$1/2$ particles. Similarly, we encode the entire quantum
circuit model as a part of the factorisable scattering model and are also able to perform universal
quantum computation by only using the two-body scattering matrix $Y^{12}_{12}$. 

 Let us conclude by making remarks on the comparison between {\em integrable quantum computing}
 \cite{ZKG05a,ZKW07,Zhang07,Zhang08} and {\em topological quantum computing} \cite{Kitaev97}. Firstly,
 the factorisation equation leads to the braid group relation with spectral parameter \cite{ZKG05a},
 and so nontrivial unitary solutions of the factorisation equation may not form a unitary braid
 representation but is able to form a unitary braid representation at special value of spectral
 parameter. Since its two-qubit quantum gates can  be chosen as a braiding gate or not,
 {\em integrable quantum computing} is viewed as  {\em topological-like} quantum computing, which is  a
 sort of hybrid quantum computing between ordinary quantum computing and {\em topological quantum
 computing}. Secondly, Kitaev's toric codes \cite{Kitaev97} form degenerate groundstates of Hamiltonian
 of the stabilizer formalism of quantum error correction codes, whereas quantum error correction codes,
 for example, the Shor nine-qubit codes, can be determined by the unitary evolution of a Hamiltonian
 associated with the factorisation equation \cite{Zhang08}.
 Thirdly, {\em topological quantum computing} is fault-tolerant due to nontrivial topological aspects
 such as nontrivial boundary conditions, whereas the fault-tolerance of {\em integrable quantum computing}
 is promised by the integrability (for example, scattering without diffractions \cite{Sutherland04}) of the
 associated quantum many-body system.

\section*{Acknowledgements}

The author wishes to thank Professor Lu Yu and Institute of Physics,
Chinese Academy of Sciences, for their hospitality and support during
the visit in which part of this work was done.

 \end{document}